\documentclass[aps,preprint,amsmath,amssymb]{revtex4-1}
\usepackage{graphicx,color}
\newcommand{\nn}{\nonumber}
\newcommand{\bd}{\begin{document}}
\newcommand{\ed}{\end{document}}
\newcommand{\bc}{\begin{center}}
\newcommand{\ec}{\end{center}}
\newcommand{\be}{\begin{eqnarray}}
\newcommand{\ee}{\end{eqnarray}}
\newcommand{\ba}{\begin{array}}
\newcommand{\ea}{\ed{array}}
\newcommand{\strich}[1]{#1  \! \! \slash}
\newcommand{\eqn}{\global\def\theequation}
\newcommand{\sw}{sin^2 \theta_W}
\newcommand{\fbd}{f_B}
\renewcommand{\thefootnote}{\alph{footnote}}
\newcommand{\se}{\section}
\newcommand{\sse}{\subsection}
\newcommand{\bi}{\bibitem}
\def\figcap{\section*{Figure Captions\markboth
     {FIGURECAPTIONS}{FIGURECAPTIONS}}\list
     {Figure \arabic{enumi}:\hfill}{\settowidth\labelwidth{Figure 999:}
     \leftmargin\labelwidth
     \advance\leftmargin\labelsep\usecounter{enumi}}}
\let\endfigcap\endlist \relax
\def\reflist{\section*{References\markboth
     {REFLIST}{REFLIST}}\list
     {[\arabic{enumi}]\hfill}{\settowidth\labelwidth{[999]}
     \leftmargin\labelwidth
     \advance\leftmargin\labelsep\usecounter{enumi}}}
\let\endreflist\endlist \relax

\def\Journal#1#2#3#4{{#1} {{\bf #2},} {#4} {(#3)}}
\def\NCA{Nuovo Cimento}
\def\NIM{Nucl. Instrum. Methods}
\def\NIMA{{Nucl. Instrum. Methods} A}
\def\NP{{Nucl. Phys.} }
\def\NPB{{Nucl. Phys.} B }
\def\NPA{{Nucl. Phys. A}}
\def\PLB{{Phys. Lett.}  B}
\def\PL{{Phys. Lett.}}
\def\PPSA{{Proc. Phys. Soc.} A}
\def\PRP{{ Phys. Rep.}}
\def\PRL{ Phys. Rev. Lett.}
\def\PR{{Phys. Rev.}}
\def\PRD{{Phys. Rev.} D}
\def\PRC{{Phys. Rev.} C}
\def\ZP{{Z. Phys.}}
\def\ZPC{{Z. Phys. C}}
\def\EPJ{{Eur. Phys. J.}}
\def\EPJC{{Eur. Phys. J.} C}
\def\ZPA{{Z. Phys.} A}
\def\MPL{{Mod. Phys. Lett.}}
\def\MPLA{{Mod. Phys. Lett.} A}
\def\CPC{Comput. Phys. Commun.}
\def\JHEP{{J. High Energy Phys.}}
\def\JPG{{J. Phys. G.}}
\def\SJNP{Sov. J. Nucl. Phys.}
\def\NCA{ Nuovo Cimento}
\def\NIM{ Nucl. Instrum. Methods}
\def\NIMA{{ Nucl. Instrum. Methods} A}
\def\NP{{ Nucl. Phys.}}
\def\ANP{{Adv. Nucl. Phys.}}
\def\CPC{{Comput. Phys. Commun.}}

\begin{document}
\title
{\Large {\bf  Semi-dileptonic decays of the  light vector mesons
in Light Front Quark Model}
}

\author{
Chao-Qiang Geng$^{1,3,4}$\footnote{E-mail address: geng@phys.nthu.edu.tw}
and
Chong-Chung Lih$^{2,4}$\footnote{E-mail address: cclih@phys.nthu.edu.tw} 
}
\affiliation{
$^1$College of Mathematics \& Physics, Chongqing University of Posts \& Telecommunications, Chongqing, 400065, China
\\
$^2$Department of Optometry, Shu-Zen College of Medicine and Management,
Kaohsiung Hsien,Taiwan 452 
\\
$^3$Department of Physics, National Tsing Hua University, Hsinchu, Taiwan 300
\\
$^4$Physics Division, National Center for Theoretical Sciences, Hsinchu, Taiwan 300
}

\date{\today}

\begin{abstract}
We study the transition form factors of the light vector to pseudoscalar mesons 
as functions of the momentum transfer $q^2$ within the light-front quark model. 
With these form factors, we calculate the decay branching ratios of all possible modes for
$V\to P\ell^+\ell^-$ ($V=\omega$ and $\phi$, $P=\pi^0$, $\eta$ and $\eta^{\prime}$ and $\ell=e$ and $\mu$).
We find that our numerical results fit with the data, such as those 
of $\omega \to \pi^0 \ell^+\ell^-$ and $\phi\to \pi^0 e^+e^-$ by NA60 and
$\phi \to\eta e^+e^-$ by SND. 
We also predict that the branching ratios of
$\phi \to \pi^0 \mu^+\mu^-$, $\omega\to \eta e^+e^-$, $\omega\to \eta \mu^+\mu^-$, 
$\phi\to \eta \mu^+\mu^-$ and $\phi\to \eta^{\prime} e^+e^-$ to be aroud
$3.48\times 10^{-6}$, $3.22\times 10^{-6}$, $1.81\times 10^{-9}$, 
$6.86\times 10^{-6}$ $2.97\times 10^{-7}$, 
respectively.
\end{abstract}

\maketitle %

\se{Introduction}

The investigations of the  transition processes between light vectors $(V=\omega$ and  $\phi)$ and pseudoscalars 
$(P=\pi^0$, $\eta$ and  $\eta^{\prime})$ mesons  are helpful to infer the internal physics properties of the mesons, 
in particular, the non-perturbative QCD effects. 
These processes  can be described by transition form factors,
denoted as $f_{V\to P}$, from the parametrizations of the hadronic matrix elements. 
Some non-perturbative QCD approaches are available to evaluated these elements, 
such as the lattice QCD, QCD sum rule and vector mesons dominant (VMD). Phenomenologically, the 
relativistic light front quark model (LFQM) also provides a convenient method to study these form factors. 

Motivated by the recent accurate data on $\omega \to \pi^0 \mu^+\mu^-$ by 
 the NA60 collaboration~\cite{na60},
we would like to study the transition form factors of $f_{V\to P}$ within the framework of the LFQM.
In particular, we concentrate on the decay processes of $V \to P \ell^+\ell^-$ 
with $V=(\omega, \phi$), $P=(\pi^0, \eta, \eta^{\prime}$ and $\ell=(e, \mu)$. 
In the LFQM, one can calculate the form factors in the frame where the
momentum transfer is purely longitudinal, $i.e$, $p_{\bot}=0$ and
$p^2=p^+p^-$, which covers the whole allowed kinematic region of $0\leq p^{2}\leq
p_{\max }^{2}$. 
We will use the phenomenological light front (LF) meson wave functions~\cite{lfqm1,vex1,lf5} 
to evaluate $f_{V\to P}$ in the LFQM. 
The LF wave functions can be constructed by the simplest structures of 
the meson constituents in terms of quark-antiquark ($Q\bar{Q}$) Fock states~\cite{lf5}, 
which have been widely applied to study the form factors 
of the Dalitz decays~\cite{lfqm1,lf2,lf5}.

The paper is organized as follows:  In Sec.~II, we present our formalism for
the transition form factors of $V\to P$ within the LFQM and the decay branching ratios of $V\to P \ell^+\ell^-$. 
In Sec.~III, we perform our numerical calculations on these processes. 
We will also compare our results in the LFQM with the data as well as other theoretical predictions.
We give our conclusions in Sec.~IV.

\se{Formalism}

We start with the decay process
\be 
\label{eq1}
V(p_v)\to P(p_p) \gamma^* (q)\to P(p_p) \ell^+(p_1)\ell^-(p_2)\,,
\ee
where $V$ represents the vector meson of $\omega$ or $\phi$, $P$ stands for the pseudoscalar of $\pi^0$ or $\eta$ or 
$\eta'$ and $\ell=e$ or $\mu$, while $p_v$, $p_p$, $q$, $p_1$ and $p_2$ are the corresponding momenta. 
The reason that we only consider the process in Eq.~(\ref{eq1}) is that the dominant contributions to 
the dilepton channels of the vector mesons are through the exchanges of the virtual photon.
The decay amplitude for the process in Eq.~(\ref{eq1}) is given by~\cite{width1}:
\be 
A
=ie^{2}f_{V\to P}(q^2)~\varepsilon_{\mu\nu\rho\sigma}\epsilon^\mu
p_{p}^\nu q^\rho\frac{1}{q^2}\,\bar{u}(p_2)\gamma^{\sigma}v(p_1)\,, \label{def}
\ee
where $q=p_1+p_2$, $\epsilon^\mu$ is the polarization vector of the vector meson, and $f_{V\to P}(q^2)$
is the transition form factor,
defined by
\be
\langle P(p_{p})|J^{em}_{\sigma}|V(p_{v})\rangle=
f_{V\to P}(q^2)~\varepsilon_{\mu\nu\rho\sigma}\epsilon^\mu
p_{p}^\nu q^\rho\,.
\ee
To calculate the form factor of $f_{V\to P}(q^2)$, we  use the quark-flavor mixing scheme 
to express the vector mesons of $\phi$ and $\omega$ as 
two orthogonal flavor states of $|\psi_{q}\rangle$ and $|\psi_{s}\rangle$ 
with one mixing angle scenario, 
parameterized as~\cite{vang,vang2}
\be
\left(
\begin{array}{c}
|\phi\rangle  \\
|\omega\rangle
\end{array}
\right)\,=
\left(
\begin{array}{cc}
\cos\theta^V & -\sin\theta^V  \\
\sin\theta^V & \cos\theta^V
\end{array}
\right)\,
\left(
\begin{array}{c}
|\psi_{q}\rangle  \\
|\psi_{s}\rangle
\end{array}
\right)\,,
\label{mix}
\ee
where $|\psi_{q}\rangle=\frac{1}{\sqrt{2}}|u\bar{u}+d\bar{d}\rangle$,
 $|\psi_{s}\rangle=|s\bar{s}\rangle$, and $\theta^V$ is the mixing angle, which has been extensively studied in
 the literature~\cite{vang}. 
In this paper, we adopt its value to be $\theta^V\simeq -3.18^{\circ}$ 
in the quark-flavor basis.  
Under this scheme, the physical states of $\omega$ and $\phi$
can be written as combinations of $Q\bar{Q}$ states:
\be
|\phi\rangle&=&\frac{\cos\theta^V}{\sqrt{2}}|u\bar{u}+d\bar{d}\rangle-
\sin\theta^V|s\bar{s}\rangle\,,\nonumber \\
|\omega\rangle&=&\frac{\sin\theta^V}{\sqrt{2}}|u\bar{u}+d\bar{d}\rangle+
\cos\theta^V|s\bar{s}\rangle\,.
\ee
For the outgoing pesudoscalar mesons, 
we use $|\pi^0\rangle=|u\bar{u}-d\bar{d}\rangle/\sqrt{2}$, while 
the states of $\eta$ and $\eta'$ can be written in terms of 
the two orthogonal states of $|\eta_{q}\rangle$ and $|\eta_{s}\rangle$ in the quark-flavor mixing scheme,
given by~\cite{phi0,phi3}
\be
\left(
\begin{array}{c}
|\eta\rangle  \\
|\eta'\rangle
\end{array}
\right)\,=
\left(
\begin{array}{cc}
\cos\beta & -\sin\beta   \\
\sin\beta &  \cos\beta
\end{array}
\right)\,
\left(
\begin{array}{c}
|\eta_{q}\rangle  \\
|\eta_{s}\rangle
\end{array}
\right)\,,
\label{mix}
\ee
where $|\eta_{q}\rangle=|u\bar{u}+d\bar{d}\rangle/\sqrt{2}$,
 $|\eta_{s}\rangle=|s\bar{s}\rangle$, and $\beta$ is the mixing angle
constrained to be 
$\beta\simeq 37^{\circ} \sim 42^{\circ}$~\cite{phi3}.
Consequently, the valence states of $\eta^{(\prime)}$
can be presented as:
\be
|\eta\rangle&=&\cos\beta\frac{|u\bar{u}+d\bar{d}\rangle}{\sqrt{2}}-
\sin\beta|s\bar{s}\rangle\,,\nonumber \\
|\eta'\rangle&=&\sin\beta\frac{|u\bar{u}+d\bar{d}\rangle}{\sqrt{2}}+
\cos\beta|s\bar{s}\rangle\,.
\ee
Combining with Eqs.~(1), (3) and (5), the transition form factor of $V\to \pi^0$ can be found by 
summing up the relevant Fock states to be 
\be
f_{V \to \pi^0}(q^2) = \frac{\cos\theta^V}{\sqrt{2}} f_{|\psi_{q}\rangle \to |\pi^0\rangle}(q^2) 
= \frac{\cos\theta^V}{\sqrt{2}} f_{V_{q\bar{q}} \to \pi^{0}}(q^2)  \,.
\label{realff1}
\ee
Similarly,
the transition from factors of $V \to \eta, \eta'$ have the forms
\be
f_{\phi\to\eta} &=& \cos\theta^V\cos\beta f_{|\psi_{q}\rangle \to |\eta_{q}\rangle} 
+ \sin\theta^V\sin\beta f_{|\psi_{s}\rangle \to |\eta_{s}\rangle}\,, \nn \\
f_{\phi\to\eta'} &=& \cos\theta^V\sin\beta f_{|\psi_{q}\rangle \to |\eta_{q}\rangle} 
- \sin\theta^V\cos\beta f_{|\psi_{s}\rangle \to |\eta_{s}\rangle}\,, \nn \\
f_{\omega\to\eta} &=& \sin\theta^V\cos\beta f_{|\psi_{q}\rangle \to |\eta_{q}\rangle} 
- \cos\theta^V\sin\beta f_{|\psi_{s}\rangle \to |\eta_{s}\rangle}\,, \nn \\
f_{\omega\to\eta'} &=& \sin\theta^V\sin\beta f_{|\psi_{q}\rangle \to |\eta_{q}\rangle} 
+ \cos\theta^V\cos\beta f_{|\psi_{s}\rangle \to |\eta_{s}\rangle}\,,
\label{mix}
\ee
where $f_{|\psi_{q}\rangle \to |\eta_{q}\rangle}$ and $f_{|\psi_{s}\rangle \to |\eta_{s}\rangle}$ 
can be replaced by $f_{V_{Q\bar{Q}} \to P_{Q\bar{Q}}}$$(Q=u, d, s)$. 

In the LFQM, a neutral meson wave function is constructed by the simple structure of  $Q\bar{Q}$ 
in terms of its constituent quark $Q$ and  anti-quark $\bar{Q}$ with the total momentum $p$ and spin $S$ as~\cite{lf5},
\begin{eqnarray}
|M(p,S,S_z)\rangle &=& \int [dk_{1}][dk_{2}] 2(2\pi)^{3}\delta
^{3}(p-k_{1}-k_{2}) \nonumber \\ &\times&\sum_{\lambda
_{1}\lambda_{2}} \Phi_{M}(k_1,k_2,\lambda_1,\lambda_2)
b_{Q}^{+}(k_{1},\lambda _{1}) d_{\bar{Q}}^{+}(
k_{2},\lambda _{2}) |\,0\,\rangle\,, \label{mwf}
\end{eqnarray}
where
\be
 [dk] = {dk^+d^{2}k_{\bot}\over 2(2\pi)^3 }\,,
\ee
 $\Phi_{M}$ is the amplitude of the corresponding $Q\bar{Q}$ and $k_{1(2)}$ 
($\lambda_{1(2)}$) is the on-mass shell LF momentum (helicity) of the internal quark.
In the momentum space, the wave function $\Phi_{M}$ 
can be expressed as a covariant form~\cite{lfqm1,vex1}
\be
\Phi_{M}(z,k_{\bot })&=&\left( \frac{%
k_{1}^{+}k_{2}^{+}}{2[M_{0}^{2}-\left( m_{Q}-m_{\bar{Q}} \right) ^{2}]}\right)^{%
\frac{1}{2}}\overline{u}\left( k_{1}, \lambda _{1}\right)
\Gamma v\left( k_{2},\lambda _{2}\right) \phi_{M}(z,k_{\bot}) \,,
 \nn \\
M_0^2&=&{ m_{\bar{Q}}^2+k_\bot^2\over z}+{ m_{Q}^2+k_\bot^2\over
1-z}\, ,
\label{n6}
\ee
where $\Gamma $ stands for
\be
&&\Gamma=\gamma_5 \qquad ({\rm pseudoscalar}, S=0),
\nonumber\\
&&\Gamma=-\not{\! \hat{\varepsilon}}(S_z)+
          {\hat{\varepsilon}\cdot(k_1-k_2)
                \over M_0+m_Q+m_{\bar{Q}}} \qquad ({\rm vector}, S=1),
\ee
and
\begin{eqnarray}
        &&\hat{\varepsilon}^\mu(\pm 1) =
                \left[{2\over p^+} \vec \varepsilon_\bot (\pm 1) \cdot
                \vec p_\bot,\,0,\,\vec \varepsilon_\bot (\pm 1)\right],
                \quad \vec \varepsilon_\bot
                (\pm 1)=\mp(1,\pm i)/\sqrt{2}, \nonumber\\
        &&\hat{\varepsilon}^\mu(0)={1\over M_0}\left({-M_0^2+p_\bot^2\over
                p^+},p^+,p_\bot\right).   \label{polcom}
\end{eqnarray}
The LF relative momentum
variables $(z,k_{\bot})$ are defined by
\be
&& k^+_1=z p^+, \quad k^+_2=(1-z) p^+\,,  \nonumber \\
&& k_{1\bot}=z p_\bot-k_\bot, \quad k_{2\bot}=(1-z)
p_\bot+k_\bot\,. \ee

In principle, the momentum distribution
amplitude $\phi_{M}(z,k_\bot)$ can be obtained by solving the
LF QCD bound state equation~\cite{lfqm1}. However, before
such first-principle solutions are available, we would have to consider phenomenological ones. 
One example that has
been  widely used in the the Gaussian type wave function~\cite{lf2}, given by
\be
\phi_{M}(z,k_{\bot})=N\sqrt{\frac{1}{N_c}\frac{dk_{z}}{dz}} \exp
\left( -\frac{\vec{k}^{2}} {2\omega_{M}^{2}}\right) \,,
\label{7}
\ee
where $N = 4 ( \pi/\omega_{M}^{2})^\frac{3}{4}$, $\vec k =
(k_{\bot}, k_z)$, and $k_z$ is defined through
\be
z = {E_Q+k_z\over
E_Q + E_{\bar{Q}}} \,,~~ \ \ 1-z = {E_{\bar{Q}}-k_z \over E_Q + E_{\bar{Q}}} \, , ~~\ \
E_i = \sqrt{m_i^2 + \vec k^2} \,
\ee
by
\be
 & &
\ \ k_{z} =\left( z -\frac{1}{2}\right) M_{0}+\frac{m_{\bar{Q}}^{2}-m_{Q}^{2}}{%
2M_{0}}~\,,~~ M_0=E_Q + E_{\bar{Q}}\, .
\ee 
and $dk_z/ dz = E_Q E_{\bar{Q}}/ z(1-z) M_0$. 
From Eq.~(3), one has
\be
\langle P_{Q\bar{Q}}(p_{p})|J^{em}_{\sigma}|V_{Q\bar{Q}}(p_{v})\rangle &=&
N_{c}\int {d^4 k_3 \over{(2 \pi)^4}}
\Lambda_{P}\Bigg\{{\rm Tr}\Bigg[\left(-\not{\! \hat{\varepsilon}}+
          {\hat{\varepsilon}\cdot(k_1-k_3)
                \over M_0+m_Q+m_{\bar{Q}}}\right)  \nn \\
&&{i(-\not{\! k_3}+m_{\bar{Q}})\over{k_3^2-m^2_{\bar{Q}}+i\epsilon}}\not{\! \epsilon_{\sigma}}
       {i(\not{\!k_2}+m_Q)\over{k_2^2-m^2_Q+i\epsilon}}
\gamma_{5} {i(\not{\! k_1}+m_Q)\over{k_1^2-m^2_Q+i\epsilon}}
\Bigg] \Bigg\}
 \nn \\
&&+(\,k_{1(3)} \leftrightarrow k_{3(1)}\,,\, m_{Q} \leftrightarrow
m_{\bar{Q}}) \,,
\label{matrix}
\ee
The hadronic matrix element of the $V\to P$
transition can be newly parametrized in terms of the
initial  and final meson momenta, given by
\be
\langle P_{Q\bar{Q}}(p_{p})|J^{em}_{\sigma}|V_{Q\bar{Q}}(p_{p}+q)\rangle=
f_{V_{Q\bar{Q}}\to P_{Q\bar{Q}}}(q^2)~\varepsilon_{\mu\nu\rho\sigma}\epsilon^\mu
p_{p}^\nu q^\rho\,,
\ee
where we have used the LF momentum variables
$(x,k_{\bot})$ and worked in the frame that the transverse
momentum is purely longitudinal, $i.e.$, $q_{\bot}$ = $0$. We note
that $q^{2}=q^{+}q^{-} \geq 0$ covers the entire range of momentum
transfers. Therefore, the trace
in Eq.~(\ref{matrix}) can be easily carried out. The
form factor $f_{V_{Q\bar{Q}} \to P_{Q\bar{Q}}}$ is then found
to be
\be
f_{V_{Q\bar{Q}} \to P_{Q\bar{Q}}}(q^{2}) &=& N_{c}\int^r_0 dx\int{d^2k_\perp\over 2(2\pi)^3}
\,\frac{2x'\phi_V^*(x',k_\perp)\phi_P(x,k_\perp)}{m_Q^2+k^2_\perp} \nn \\
&\times& \Bigg\{ m_Q + 
\frac{1}{W_V}\left[rk_\perp^2+(1-r)\left(2xM_{0_{P}}k_{z}
-\frac{x'k^{2}_{\bot}}{1-x'}\right)\right]\Bigg\}\,,  
\label{fvp}
\ee
where $N_c=3$ is the number of colors and 
\be
x=r x'\,,~~~ r=\frac{m_{V}^{2}+m_{P}^{2}-q^2
+\sqrt{(m_{V}^{2}+m_{P}^{2}-q^2)^{2}-4m_{V}^{2}m_{P}^{2}}}{2m_{V}^{2}}\,.
\ee
At $q^2=0$, the form factor in Eq.~(\ref{fvp}) is evaluated to be
\be
f_{V_{Q\bar{Q}} \to P_{Q\bar{Q}}}(0) &=& N_{c}\int dx\int{d^2k_\perp\over 2(2\pi)^3}
\,\frac{2x\phi_V^*(x,k_\perp)\phi_P(x,k_\perp)}{m_Q^2+k^2_\perp}\,
 \left( m_Q + 
\frac{k_\perp^2}{W_V}\right)\,.  
\label{fvp0}
\ee
It has been noted~\cite{BCJ} that in the purely longitudinal $q^+>0$ frame, the nonvalence contribution~\cite{NV} to the form factor
has to be included and the frame dependence should be checked~\cite{Frames}.  
To avoid the nonvalence contribution,
one may use 
the Drell-Yan-West frame, $i.e.$ $q^+=0$, in which it contains only the valence part~\cite{BCJ}. In this frame, we obtain
\be
f_{V_{Q\bar{Q}} \to P_{Q\bar{Q}}}(q^{2}) &=& N_{c}\int^1_0 dx\int{d^2k_\perp\over 2(2\pi)^3}
\,\frac{2x\phi_V^*(x,k_\perp)\phi_P(x,k'_\perp)}{m_Q^2+k^2_\perp} \nn \\
&\times& \Bigg\{ m_Q +
\frac{2}{W_V}\left[k_\perp^{2}+\frac{(k_\perp q_\perp)^2}{q^2}\right]\Bigg\}\,
\label{fvpDY}
\ee
with $k_\perp=(x-1) q_\perp+k'_\perp$,
which also agrees with Eq.~(D2) in Ref.~\cite{BCJ} and Eq.~(4.13) in Ref.~\cite{jaus}, respectively. 
At $q^2=0$, the formula for $f_{V_{Q\bar{Q}} \to P_{Q\bar{Q}}}(0)$ from Eq.~(\ref{fvpDY}) is the same as that in Eq.~(\ref{fvp0}),
which is a frame independent quantity as expected.
For $q^2\neq 0$, we remake that the results in the two frames should not be different too much in our cases of
the light-to-light transitions~\cite{BCJ}.

The interaction between the photon and leptons is given by the
conventional QED~\cite{width1}. One easily obtains the differential decay rates 
normalized to the radiative decay widths of $V \to P\gamma$ as
\be
{d\,\Gamma(V \to P\ell^+ \ell^-)
\over{\Gamma(V \to P\gamma)\,dq^2}}&=&\frac{\alpha}
{3\,\pi}\frac{1}{q^2}
\,\left(1-{4\,m^2_{\ell}\over{q^2}}\right)^{1/2}\left(1+{2\,m^2_{\ell}
\over{q^2}}\right)\,    \\
&\times& 
\left[\left(1+\frac{q^2}{m_{V}^{2}-m_{P}^{2}}\right)^{2}
-\frac{4 q^2 m_{V}^{2}}{(m_{V}^{2}-m_{P}^{2})^2}\right]^{3/2}
\,\left|\frac{f_{V\to P}(q^2)}{f_{V\to P}(0)}\right|^2\,.
\label{llr}
\ee

\se{numerical results}

To numerically calculate the transition form factor of 
$f_{V\to P}$, we need to specify the quark masses of $m_{u,d,s}$ and 
the parameters appearing in $\phi_M(x,k_\bot)$ ($M=V$ or$P$). 
Here, we have chosen the quark masses as $m_q =m_{u,d}=0.25$ and
$m_s=0.42$ in GeV. To constrain the meson scale
parameter of $\omega_M$ in Eq. (\ref{7}), 
we use the branching ratio of $V \to P\gamma$, given by
\be
{\cal B}(V \to P\gamma)&=& \frac{\alpha(M_{V}^{2}-M_{P}^{2})^{3}}{24M_{V}^{3} \Gamma_V} 
|f(0)_{V \to P}|^2 \,.
\ee
Explicitly, we take~\cite{pdg}
\be
{\cal B}^{exp}(\omega \to \pi^{0}\gamma)&=&\,(8.28\pm0.28)\times 10^{-2}\,,   \\
{\cal B}^{exp}(\phi \to \pi^{0}\gamma)&=&(1.27\pm0.06)\times 10^{-3}\,. \label{br2r}
\ee
which lead to $|f_{\omega \to \pi^{0}}(0)|=(2.297\pm0.05)\times 10^{-3}$ and 
$|f_{\phi \to \pi^{0}}(0)|=(1.33\pm0.038)\times 10^{-4}$ in MeV$^{-1}$, respectively.
We can then fit the parameters of $w_{\omega, \phi}$ from Eq. (\ref{fvp0}) if we input the quark masses.

\begin{figure}[htbp]
\includegraphics*[width=4in,height=3in,angle=0]{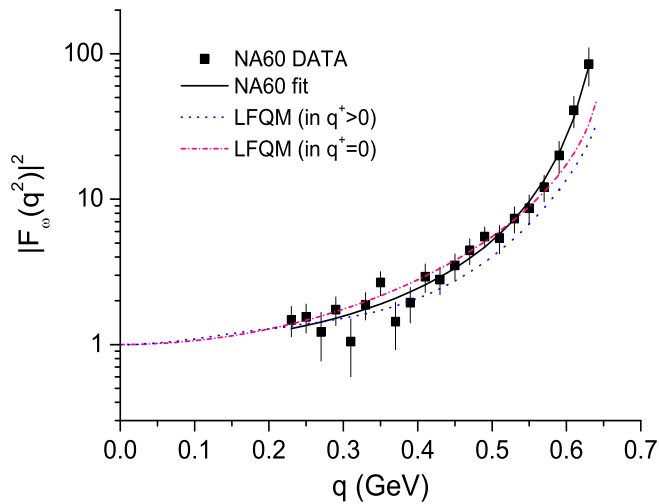}
\caption{(Color online) $F_{\omega}(q^2)\equiv f_{\omega \to \pi^{0}}(q^2)/f_{\omega \to \pi^{0}}(0)$ 
in the LFQM. }
\label{Fig1}
\end{figure}

The numerical results for $F_{\omega}(q^2)\equiv f_{\omega \to \pi^{0}}(q^2)/f_{\omega \to \pi^{0}}(0)$ 
in the $q^+>0$ and  $q^+=0$ frames of the LFQM are shown in  Fig.~\ref{Fig1}. 
From the figure, we see that our results in the LFQM, particularly the one in  $q^+=0$ fit well 
with the NA60 experimental data~\cite{na60} even though they are slightly lower for the large $q^2$ region. 
Note that the 
numerical values in the  $q^+=0$ frame  are slightly larger than those in  $q^+>0$ in  the large $q^2$ region. The difference may result from the nonvalence part in the $q^+>0$ frame~\cite{BCJ}. 
In the following calculations, we will only use the form factors evaluated in the $q^+=0$ frame.

\begin{figure}[htbp]
\includegraphics*[width=4in,height=3in,angle=0]{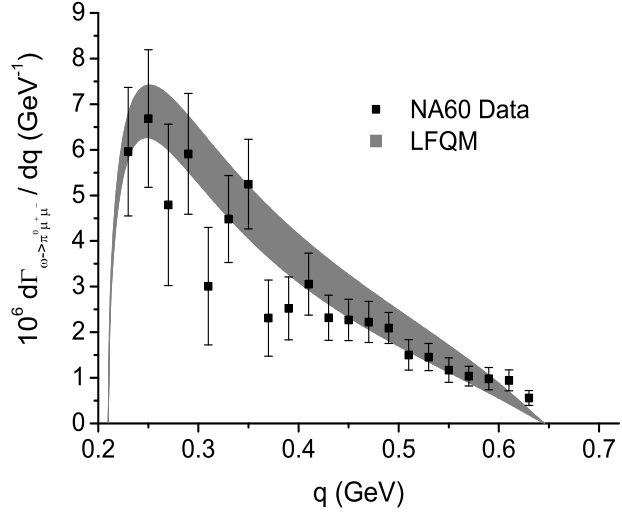}
\caption{ 
Differential decay width  of $\omega\to\pi^{0}\mu^{+} \mu^{-}$ as a function of $q$  in the LFQM,
where the small band is due to the uncertainty of the data in Eq.~(27).}
\label{Fig2}
\end{figure}\begin{figure}[htbp]
\includegraphics*[width=4in,height=3in,angle=0]{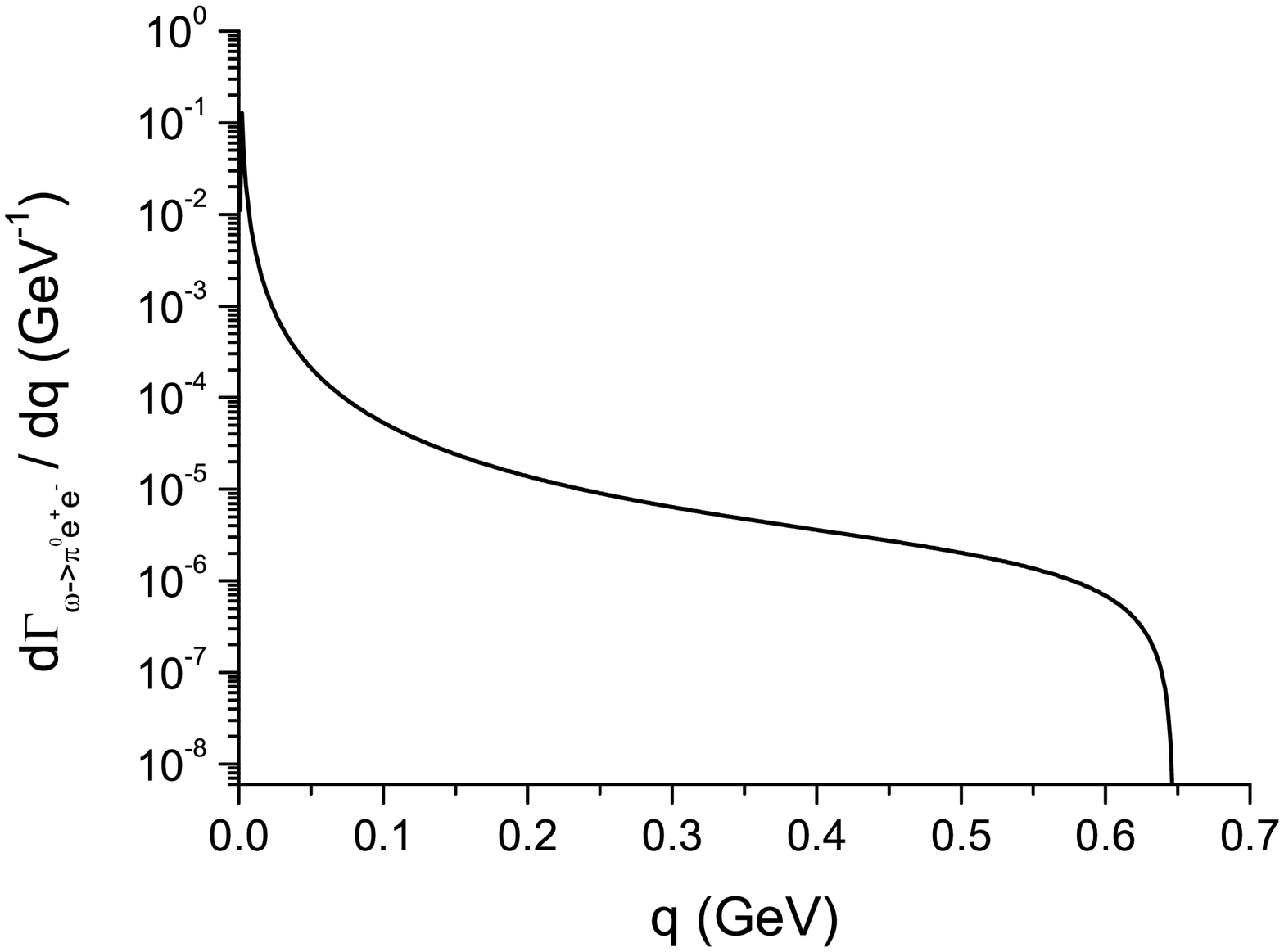}
\caption{ 
Same as Fig.~2 but for the $e$ mode.}
\label{Fig3}
\end{figure}
\begin{figure}[htbp]
\includegraphics*[width=4in,height=3in,angle=0]{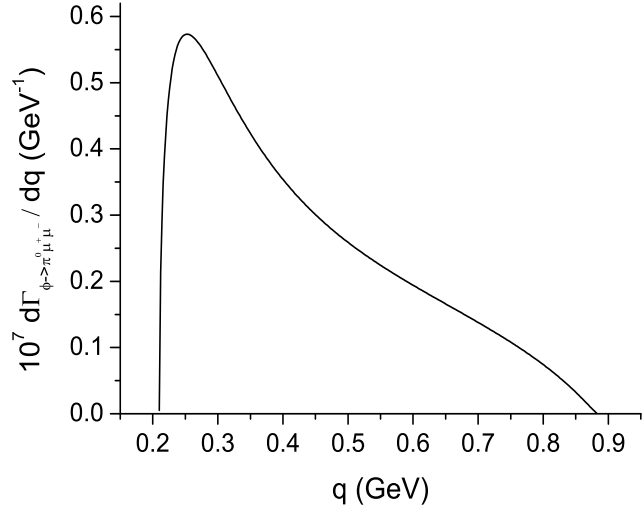}
\caption{ Differential decay width  of 
 $\phi\to\pi^{0}\mu^+ \mu^{-}$ as a function of $q$  in the LFQM.}
\label{Fig4}
\end{figure}\begin{figure}[htbp]
\includegraphics*[width=4in,height=3in,angle=0]{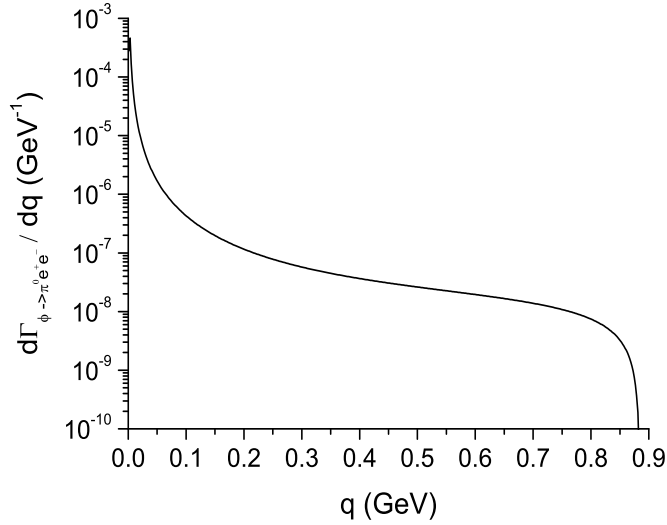}
\caption{ Same as Fig.~4 but for the $e$ mode.}
\label{Fig5}
\end{figure}

As seen from Eq. (\ref{llr}),
 the differential decay widths of $V \to P \ell^+ \ell^-$ depend on the factor of $1/q^2$, highly suppressed by the phase space. 
In Figs.~\ref{Fig2}$\sim$\ref{Fig5}, we display  the differential decay
widths of $V \to \pi^{0}\ell^+ \ell^-$ ($V=\omega$, $\phi$ and $\ell=e$, $\mu$) 
as functions of $q$ in the LFQM. Form Fig.~\ref{Fig2}, we find that our result for 
$\omega\to\pi^{0}\mu^+ \mu^{-}$ is consistent with the experimental data from NA60~\cite{na60}. 
Integrating over $q^2$ in Eq.~(\ref{llr}), we can obtain the branching ratios of $V \to \pi^0\ell^+ \ell^-$. 
Our results in the LFQM are shown in Table~\ref{Table1}. 
\begin{table}[htbp]
\caption{ Decay branching ratios  of $V \to \pi^{0}\ell^+ \ell^-$ ($V=\omega$, $\phi$ and $\ell=e$, $\mu$)
in the LFQM, where the experimental data listed in PDG~\cite{pdg} as well as the theoretical predictions 
based on the vector meson method (VMD)~\cite{vmd1} and dispersion relations (DR)~\cite{vmd2} are also given.}
\label{Table1}
\begin{tabular}{|c||c|c|c|c|} \hline
Model & LFQM & PDG~\cite{pdg} & VMD~\cite{vmd1} & DR~\cite{vmd2}
\\ \hline \hline
$10^4 {\cal B}_{\omega \to\pi^0 e^+e^-} $ & $7.82\pm 0.39$ 
& $ 7.7\pm 0.6 $ & $ 7.9 $ & $ 7.6...8.1 $
\\ \hline
$10^4 {\cal B}_{\omega \to\pi^0 \mu^+ \mu^-} $ & $ 1.21\pm0.15 $
& $ 1.3\pm 0.4 $ & $0.92 $ & $ 0.94...1.0 $
\\ \hline
$10^5 {\cal B}_{\phi \to\pi^0 e^+e^-} $ & $1.35\pm0.12$
& $ 1.12\pm 0.28 $ & $1.6 $ & $ 1.39...1.51 $
\\ \hline
$10^6 {\cal B}_{\phi \to\pi^0 \mu^+ \mu^-} $ &$ 3.48\pm0.57 $ 
& $ - $ & $4.8 $ & $ 3.7...4.0 $
\\ \hline
\end{tabular}
\end{table}
 In the table, we also give
the experimental data listed in PDG~\cite{pdg} as well as the theoretical predictions based on the vector meson method (VMD)~\cite{vmd1} 
and dispersion relations (DR)~\cite{vmd2}.
 From  Table~\ref{Table1}, we observe that our predicted values for $\omega \to\pi^0 e^+e^-$ and $\pi^0 \mu^+\mu^-$
in the LFQM agree with those in  PDG~\cite{pdg} in errors 
and also are  close to the theoretical values in VMD~\cite{vmd1} 
and DR~\cite{vmd2}, respectively.
On the other hand, 
our results for the $\phi \to\pi^0 \ell^+ \ell^-$ decay modes are slightly smaller than those in Refs.~\cite{vmd1,vmd2}. 
However, it is interesting to note that the decay branching ratio 
of $\phi \to\pi^0 e^+e^-$ in the LFQM is consistent  with the experimental data~\cite{pdg}.

\begin{figure}[htbp]
\includegraphics*[width=4in,height=3in,angle=0]{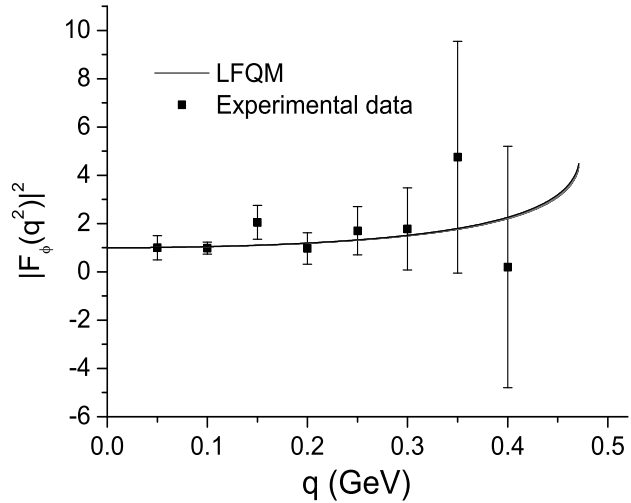}
\caption{ 
$F_{\phi}(q^2)\equiv f_{\phi \to \eta}(q^2)/f_{\phi \to \eta}(0)$ in the LFQM.}
\label{Fig6}
\end{figure}

We now consider the processes of $V \to \eta^{(\prime)}$ ($V=\omega$, $\phi$).
 In Fig.~\ref{Fig6},  we show the transition form factor of 
 $F_{\phi}\equiv f_{\phi \to \eta}(q^2)/f_{\phi \to \eta}(0)$ in the LFQM.
In the figure, we also plot the experimental data from  SND  at the VEPP-2M collider~\cite{snd}. 
Our result is consistent with the current data. 
Since the data contain large errors when $q > 0.2$ GeV, more precise 
experimental measurements are clearly needed to test our model. 
In  Table~\ref{Table2}, 
\begin{table}[htbp]
\caption{ Decay branching ratios  of $V \to \eta^{(')}\ell^+ \ell^-$ ($V=\omega$, $\phi$ and $\l=e$, $\mu$),
where we have also shown the experimental data in PDG~\cite{pdg} and the theoretical results 
based on the  VMD~\cite{vmd1} and large $N_c$~\cite{vmd3} calculations.}
 \vskip 0.2in
\label{Table2}
\begin{tabular}{|c||c|c|c|c|} \hline
Model & LFQM & PDG~\cite{pdg} & VMD~\cite{vmd1} & Large $N_{c}$~\cite{vmd3}
\\ \hline \hline
$10^6 {\cal B}_{\omega \to\eta e^+e^-} $ & $3.22\pm 0.28$ 
& $ - $ & $ 6.0 $ & $ 3.20\pm0.10 $
\\ \hline
$10^{9} {\cal B}_{\omega \to\eta \mu^+ \mu^-} $ & $ 1.81\pm0.23 $
& $ - $ & $ 1.8 $ & $ 1.00\pm0.00 $
\\ \hline
$10^4 {\cal B}_{\phi \to\eta e^+e^-} $ & $1.07\pm0.02$
& $ 1.15\pm 0.10 $ & $1.1 $ & $ 1.09\pm0.06 $
\\ \hline
$10^6 {\cal B}_{\phi \to\eta \mu^+ \mu^-} $ &$ 6.86\pm0.18 $ 
& $ < ~9.4 $ & $6.8 $ & $ 6.44\pm0.69 $
\\ \hline
$10^7 {\cal B}_{\phi \to\eta' e^+ e^-} $ &$ 2.97\pm0.10 $ 
& $ - $ & $ - $ & $ - $
\\ \hline
\end{tabular}
\end{table}
we list all possible decay branching ratios of  $V \to \eta^{(')}\ell^+ \ell^-$ in the LFQM.
In the table, we have also shown the experimental data in PDG~\cite{pdg} and the theoretical results 
based on the  VMD~\cite{vmd1} and large $N_c$~\cite{vmd3} calculations.
Note that most of these decay modes have not been measured yet
except $\phi \to\eta e^+e^-$. From the table, we see that our results are close to those 
in the large $N_c$ calculations~\cite{vmd3} except the one for $\omega\to \eta \mu^+\mu^-$. 
The prediction of  $\omega \to\eta e^+e^-$ 
in the VMD model is about two times  larger than ours  but those of $\omega\to \eta \mu^+\mu^-$ and
$\phi \to\eta \ell^+\ell^-$ are consistent with the ones in the LFQM. 
Comparing with the only experimental value
in $\phi \to\eta e^+e^-$, our result fits well  with the data in error.

\se{Conclusions}

We have studied the form factors of the light vector mesons ($\omega$ and $\phi$)
 into  pseudoscalar mesons ($\pi^0$, $\eta$ and $\eta^\prime$) in the LFQM. 
 In our calculations, we have adopted the Gaussian-type wave function and evaluated 
 the form factors for the momentum dependences  in all allowed $q^2$ regions. 
Our numerical results of $f_{\omega \to \pi^{0}}(q^2)/f_{\omega \to \pi^{0}}(0)$ are
 close to the experimental data by NA60.
 Similarly, our values for that of  $\phi \to\eta e^+e^-$ compare well with
 the  data by SND  at the VEPP-2M collider.
With these form factors, we have calculated all possible decay branching ratios of
$V\to P\ell^+\ell^-$  with $V=\omega$ and $\phi$, $P=\pi^0$, $\eta$ and $\eta^{\prime}$ and $\ell=e$ and $\mu$.
Explicitly, we have found that our numerical results can fit with the data, 
such as those for $\omega \to \pi^0 \ell^+\ell^-$ and $\phi\to \pi^0 e^+e^-$
by NA60 and $\phi \to\eta e^+e^-$ by SND. 
We also predict that the branching ratios of
$\phi \to \pi^0 \mu^+\mu^-$, $\omega\to \eta e^+e^-$, $\omega\to \eta \mu^+\mu^-$, 
$\phi\to \eta \mu^+\mu^-$, and $\phi\to \eta^{\prime} e^+e^-$
to be
$(3.48\pm0.57)\times 10^{-6}$, $(3.22\pm0.28)\times 10^{-6}$, $(1.81\pm0.23)\times 10^{-9}$, 
$(6.86\pm0.18)\times 10^{-6}$ $(2.97\pm0.10)\times 10^{-7}$
 respectively. Some of these modes could be measured in the future experiments.

\section{Acknowledgments}
 This work was partially supported by National Center for Theoretical
Sciences, SZL10204006, National Science Council  
 (NSC-102-2112-M-471-001-MY3 and
 NSC-101-2112-M-007-006-MY3) and National Tsing Hua
University (102N2725E1).\\


\begin{thebibliography}{99}

\bi{na60}
R. Arnaldi $et\ al.$ [NA60 Collaboration], Phys. Lett {\bf B667}, 206 (2009);
G. Usai $et\ al.$ [NA60 Collaboration], Nucl. phys. {\bf A855}, 189 (2011).

\bi{lfqm1}
K. G. Wilson, T. S. Walhout, A. Harindranath, W. M. Zhang, R. J. Perry and S. D. Glazek, 
Phys. Rev. {\bf D49}, 6720 (1994);
C.~Q.~Geng, C.~C.~Lih and W.~M.~Zhang, Phys. Rev. {\bf D57}, 5697 (1998);
 Phys. Rev. {\bf D62}, 074017 (2000).

\bi{vex1}
W. Jaus, Phys. Rev. {\bf D41} (1990) 3394; {\bf 44} (1991) 2851.

\bi{lf5}
C.~C.~Lih and C.~Q.~Geng, Phys. Rev. {\bf C85}, 018201 (2012).


\bi{lf2}
C.~Q.~Geng, C.~C.~Lih and W.~M.~Zhang, 
 Mod. Phys. Lett. {\bf A15} (2000) 2087;
C.~C.~Lih, C.~Q.~Geng and W.~M.~Zhang, Phys. Rev. {\bf D59}, 114002 (1999);
C. Q. Geng, C. C. Lih and C. C. Liu, Phys. Rev. {\bf D62}, 034019 (2000);
C. H. Chen, C. Q. Geng, C. C. Lih and C. C. Liu, Phys. Rev. {\bf D75}, 074010 (2007).


\bi{width1}
L. G. Landsberg, Phys. Rept. {\bf 128}, 301 (1985).

\bi{vang}
T. Ohshima, Phys. Rev. {\bf D22} 707 (1980); 
R. Escribano and J.-M. Frere, JEPT {\bf 0506} 209 (2005); 
A. Bramon, R. Escribano and M. D. Scadron, Phys. Lett. {\bf B403} 339 (1997); 
M. Benayoun. L. DelBuono, Ph. Leruste and H. B. O'Connell, Eur. Phys. J. {\bf C17} 303 (2000); 
W. Qian and B. Q. Ma, Phys. Rev. {\bf D78} 074002 (2008).

\bi{vang2}
F. Giacosa and G. Pagliara, Nucl. Phys. {\bf A833} 138-155 (2010).

\bi{phi0}
 T.~Feldmann, P.~Kroll and B.~Stech,
 Phys.\ Rev.\ D {\bf 58}, 114006 (1998);
 T.~Feldmann, P.~Kroll and B.~Stech,
 Phys.\ Lett.\ B {\bf 449}, 339 (1999);
 T.~Feldmann,
 Int.\ J.\ Mod.\ Phys.\ A {\bf 15}, 159 (2000);
 A.~Bramon, R.~Escribano and M.~D.~Scadron,
 Eur.\ Phys.\ J.\ C {\bf 7}, 271 (1999);
P. Kroll, Mod. Phys. Lett. {\bf A20}, 2667 (2005);
F. G. Cao and A.I. Signal, Phys. Rev. {\bf D85} 114012 (2012).

\bi{phi3}
F. G. Cao, Phys. Rev. {\bf D85} 057501 (2012); C. E. Thomas, JHEP {\bf 0710} 026 (2007); 
F. Ambrosino $et\ al.$, JHEP {\bf 0907} 105 (2009); 
H. W. Ke, X. H. Yuan and X. Q. Li, Int. J. Mod. Phys. {\bf A26} 4731 (2011). 

\bi{BCJ}
B. L.G. Bakker, H. M. Choi and C. R. Ji, Phys. Rev. {\bf D67} 113007 (2003).

\bi{NV}
C.~R.~Ji and H.~M.~Choi,
  Phys.\ Lett.\ B {\bf 513}, 330 (2001);
B.~L.~G.~Bakker, H.~M.~Choi and C.~R.~Ji,
  Phys.\ Rev.\ D {\bf 65}, 116001 (2002);
   H.~M.~Choi and C.~R.~Ji,
  Phys.\ Rev.\ D {\bf 72}, 013004 (2005).
  
\bi{Frames}
 C.~R.~Ji and C.~Mitchell,
  Phys.\ Rev.\ D {\bf 62}, 085020 (2000);
  B.~L.~G.~Bakker and C.~R.~Ji,
  Phys.\ Rev.\ D {\bf 65}, 073002 (2002);
  H.~M.~Choi and C.~R.~Ji,
  arXiv:1311.0552 [hep-ph].
  
\bi{jaus}
Wolfgang Jaus, Phys. Rev. {\bf D60} 054026 (1999).

\bi{pdg}
J. Beringer $et\ al.$ [Particle Data Group], Phys. Rev. {\bf D86}, 010001 (2012).

\bi{vmd1}
A. Faessler, C. Fuchs and M.I. Krivoruchenko, Phys. Rev. {\bf C61}, 035206 (2000).

\bi{vmd2}
S. P. Schneider, B. Kubis and F. Niecknig, Phys. Rev. {\bf D86} 054013 (2012).

\bi{snd}
M. N. Achasov, $et\ al.$,  Phys. Lett. {\bf B504}, 275 (2001).

\bi{vmd3}
 C.~Terschlusen and S.~Leupold,
  Phys.\ Lett.\ {\bf B691}, 191 (2010).

\end{thebibliography}
\end{document}